\documentclass[prl,aps,twocolumn,floatfix,10pt,amssymb]{revtex4} 

\usepackage{graphicx} 
\usepackage{amsmath}
\usepackage{mciteplus}
\usepackage{natbib}

\begin{document} 

\title{Integer partition manifolds and phonon damping in one dimension}   

\author{I. E. Mazets$^{1,2}$ and N. J. Mauser$^2$} 
\affiliation{$^1$Vienna Center for Quantum Science and Technology, Atominstitut, TU~Wien,~Stadionallee~2, A-1020~Vienna,Austria\\
$^2$ Wolfgang Pauli Institute c/o Fakult\"at f\"ur Mathematik, Universit\"at Wien, Oskar Morgenstern Platz 1, A-1090 Vienna, Austria }
\begin{abstract} 
We develop a quantum model based on the correspondence between energy distribution between harmonic oscillators and the 
partition of an integer number. A proper choice of the interaction Hamiltonian acting within this manifold of states 
allows us to examine both the quantum typicality and the non-exponential relaxation in the same system. A quantitative agreement 
between the field-theoretical calculations and the exact diagonalization of the Hamiltonian is demonstrated. 
\end{abstract} 
\maketitle 


The problem of thermalization in isolated quantum systems is almost as old as the quantum theory itself. 
Von Neumann pioneered in this field with his proof of the quantum ergodic theorem 
\cite{Neumann1929} (see also the English translation \cite{von2010proof} and an extended commentary \cite{Goldstein2010}). 
Another important  achievement in understanding the emergence of thermodynamics from quantum theory 
relies on the random matrix theory \cite{Wi1,Wi2,Wi3,Dyson62}, which 
is often considered as the basis for the definition of quantum chaos \cite{Bohigas86}. 
A new approach has been put forward independently by Deutsch \cite{Deutsch} and later by  
Srednicki \cite{Srednicki94}, who named it the eigenstate thermalization hypothesis (ETH). The ETH allows for quantitative 
estimations of correlations and fluctuations in thermalizing quantum systems \cite{Srednicki99}. The relation between 
von Neumann's  approach and the ETH is elucidated in Refs. \cite{RigolANDSrednicki,RigolReview}. 

Numerical tests of these theoretical concepts usually employ such models as hard-core bosons \cite{Rigol2008hardcore} 
or fermions \cite{Marquardt} on a lattice or spin chains \cite{Haque}. The problem is that the dimension of the Hilbert space 
for a half-filled lattice or a spin system with zero projection of the total spin grows exponentially with the system size. 
This is especially restrictive in the case of studying the Bose-Hubbard model with a finite interaction strength 
\cite{Haque,Altman}. A numerical method to the calculation of correlations in 
integrable systems describable by the Bethe \textit{ansatz}, i.e.,  
in Heisenberg spin chains and in one-dimensional (1D) Bose gases has been developed \cite{Caux2009}; this method is an example 
of a \textit{tour de force} in combining analytic and numerical approaches. However, the method of reducing the dimensionality of the 
Hilbert space to the states providing most of the contribution to the correlation functions in Ref. \cite{Caux2009} is restricted 
to integrable systems only.

Therefore it is always desirable to develop a model that is capable of solving experimentally relevant problems and has the 
dimension growing subexponentially with the number of the modes taken into account. In this 
Letter  
we present such a model. 
Consider a set of harmonic oscillators with the frequencies equal to an integer multiple of $\Omega $, i.e., 
$\omega _k=k\Omega $, each integer number $k=1,\, 2,\, 3,\, \dots \, $ appearing once and only once. 
Obviously, the state with the energy $n \hbar \Omega $ has a degeneracy 
equal to the number of partitions of the integer number $n$. Indeed, a partition represents $n$ as a sum of positive integers, 
the number $k$ appearing in the sum $N_k$ times:  
\begin{equation} 
\sum _k kN_k = n .  
\label{eq:1} 
\end{equation}
Obviously, the number of ways to distribute the excitation energy $n \hbar \Omega $ among 
the oscillators with frequencies $\omega _k$ is equal to the number $p(n)$ of all partitions of the integer $n$. 
Therefore we call a set of all states $|\{ N_k\} \rangle $ satisfying the condition (\ref{eq:1})
where $N_k$ is the number of quanta in the $k$th oscillator, an integer partition manifold (IPM). 

If we introduce an interaction with a typical matrix element much smaller than $\Omega $, then --- to the lowest order of the 
perturbation theory --- we can neglect mixing of different IPMs and consider coupling of levels within each IPM separately. 
Each oscillator will be regarded as a bosonic mode of a system. We see that the number $n$ characterizes the highest mode that 
can be occupied by pumping all the available energy into that mode. The dimension $D_n$ of the IPM is equal to the 
number of partitions $p(n)$ of the integer $n$. 
An asymptotic estimation for $p(n)$  was found by Hardy and Ramanujan 
100 years ago \cite{HardyRamanujan}
\begin{equation} 
p(n) \approx \frac {\exp \left( \pi \sqrt{2n/3}\right) }{4n\sqrt{3}} .
\label{eq:2}
\end{equation}
This expression grows slower than an exponential function of $n$, which makes the model suitable for exact diagonalization: 
we can take into account many modes without exploding the dimension of the Hilbert space.    

To provide a physical example where an IPM spans the Hilbert space, we consider the damping of a phonon in a 1D quasicondensate 
\cite{Andreev,Samokhin}. The $k$th mode is identified with a phonon mode with the momentum $kQ_\mathrm{min}$. To discretize the spectrum 
of phonons, as required by the construction of IPMs, we assume a finite size $L$ of the 1D system, thus introducing the smallest possible 
phonon momentum $Q_\mathrm{min}=2\pi \hbar /L$. The interaction Hamiltonian is then 
\begin{equation} 
\hat H_\mathrm{int} =\eta \sum _{k=1}^n \sum _{q=1}^{[k/2]} \sqrt{kq(k-q)} \left( \hat b_{k-q}^\dag \hat b_q^\dag \hat b_k +
\hat b_k^\dag \hat b_{k-q} \hat b_q\right) , 
\label{eq:3}
\end{equation}
where the operator $\hat b_k$ ($\hat b_k^\dag $) annihilates (creates) an excitation in the $k$th mode, 
$[k/2]$ denotes an entire part of $k/2$,  and 
$\eta = \sqrt{9\pi Q_\mathrm{min}^2\hbar c/(16\varrho _\mathrm{1D})}$, where $c$ is the speed of sound and $\varrho _\mathrm{1D}$ is the 
1D mass density of the quasicondensate \cite{Andreev,Samokhin,Bhattacherjee}. Note that this problem is different from the decay 
of a Bogoliubov quasiparticle into three quasiparticles \cite{RistivojevicMatveev} due to integrability-breaking 
effective three-body elastic collisions \cite{Mazets2008}. Numerical solution of the Gross-Pitaevskii equation \cite{Grisins2011} 
demonstrated the non-trivial dynamics of phonons in a 1D quasicondensate in the integrable limit. Therefore we find it 
interesting to elucidate the quantum origins of phonon relaxation in 1D. 

\begin{figure}[t]
\vspace*{1mm} 
\includegraphics[width=\columnwidth]{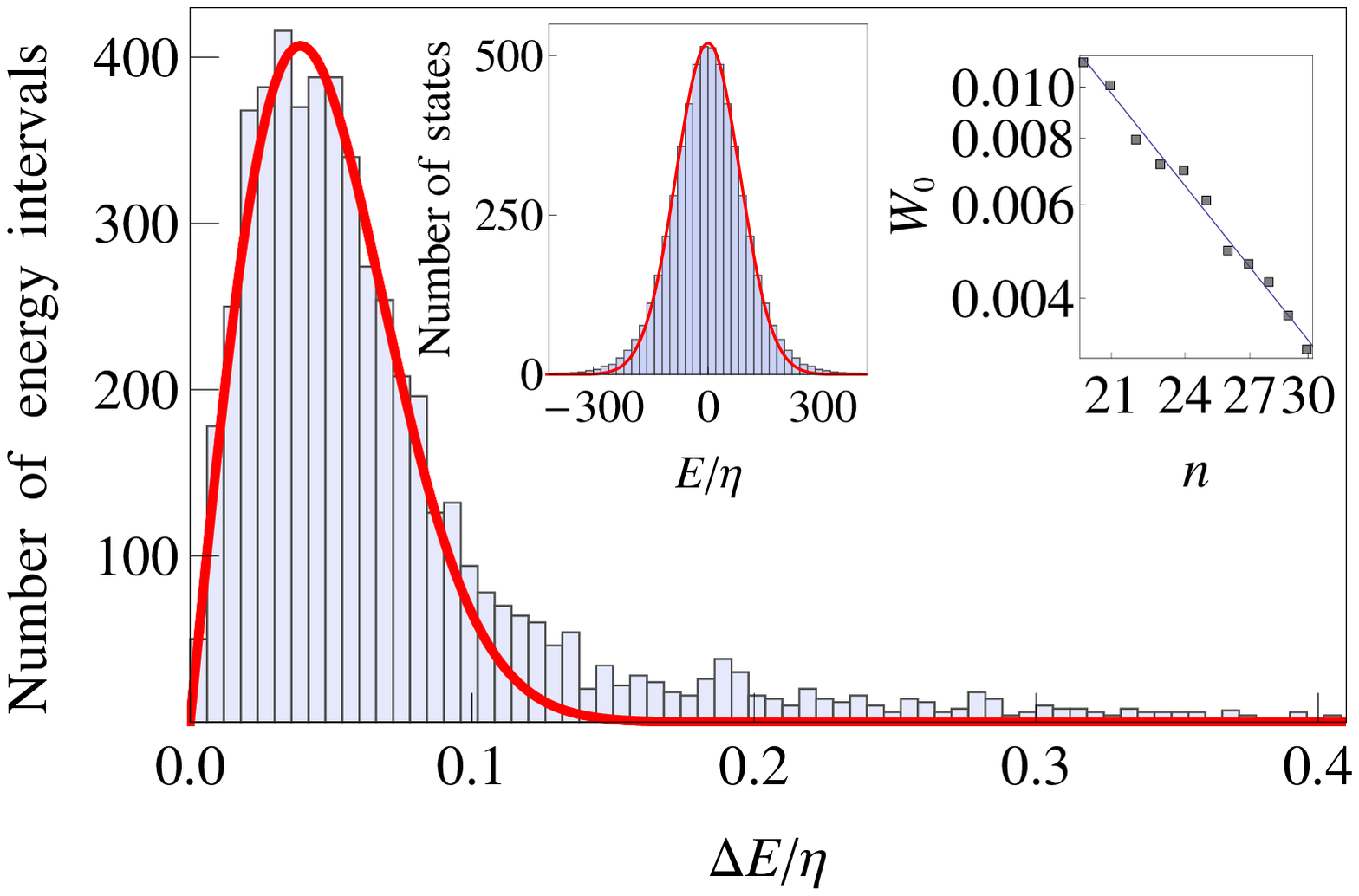} 
\begin{caption} 
{(Color online) Histogram of the energy intervals between the neighboring eigenstates of $\hat H_\mathrm{int}$ (except the states 
with $E=0$) for $n=30$. Red line: Wigner distribution. Left inset: Histogram of eigenenergies of $\hat H_\mathrm{int}$ for $n=30$. 
Red line: Gaussian distribution. Energies are scaled to $\eta $ (units on the axes of this and subsequent plots are 
dimensionless). Right inset: log-log plot of 
the fraction of states with $E=0$ in IPMs as a function of $n$. Solid line: approximation $W_0=90/n^3$.  }
\label{fig:1}
\end{caption} 
\end{figure}

We performed exact diagonalization of the Hamiltonian (\ref{eq:3}) for $n$ from 20 to 30, $\hat H_\mathrm{int}|\psi _j\rangle =
E_j|\psi _j\rangle $, using the states $|\{ N_k\} \rangle $ as the basis. The standard matrix diagonalization package of 
\textsc{Mathematica~8.0} has been applied \cite{Math8}.  
First of all, we notice that the 
distribution of eigenvalues of $\hat H_\mathrm{int}$ is close to a Gaussian (the left inset in Fig.~\ref{fig:1}) with the width 
proportional to $n^s$ with $s\approx 1.33$. A more detailed examination reveals that the eigenvalue $E=0$ is always degenerate. 
The ratio $W_0$ of the number of states with $E=0$ (within the numerical error $\sim 10^{-14}\eta $) to the dimension $D_n$ of the 
IPM can be approximated as $W_0\approx 90/n^3$ (right inset in Fig.~\ref{fig:1}). 

\begin{figure}[b] 
\vspace*{1mm} 
\includegraphics[width=\columnwidth]{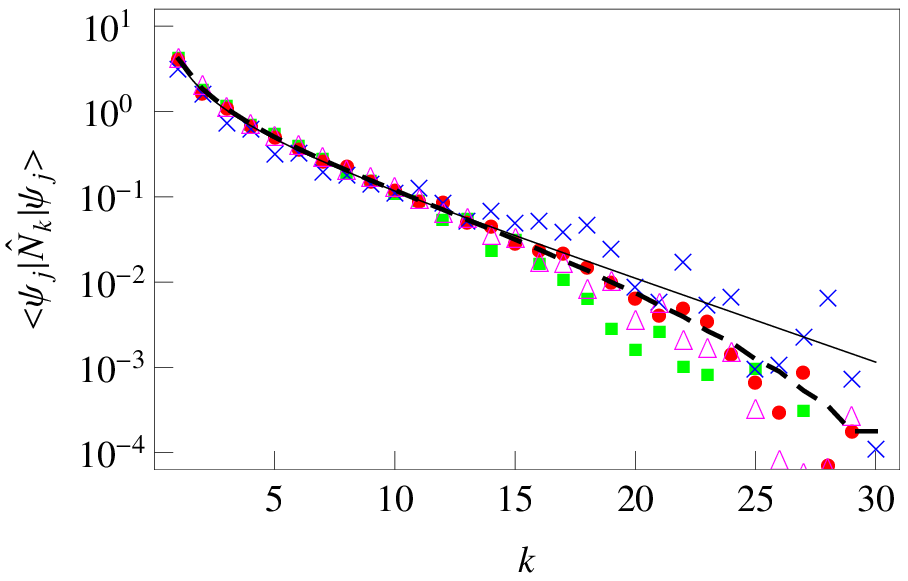} 
\begin{caption} 
{(Color online) Populations of the modes for the eigenstates with energy $E_j/\eta =$ 0 (green squares), 95.349 (red circles), 
$-190.564$ (blue crosses), 59.057 (magenta triangles); $n=30$. Solid line: the corresponding 
Bose-Einstein distribution. Dashed line: 
$\bar N_k$, see Eq. (\ref{eq:6}). The scale on the vertical axis is logarithmic.} 
\label{fig:2} 
\end{caption} 
\end{figure} 

The repulsion of levels resulting in energy intervals $\Delta E$ between eigenstates obeying the Wigner distribution 
$\propto \Delta E \exp [-(\Delta E/{\Delta  E_*})^2]$, where ${\Delta  E_*}$ sets the typical energy scale, 
is a signature of quantum chaos \cite{Bohigas86}. The main plot in Fig.~\ref{fig:1} shows the distribution of energy intervals 
between the eigenstates of $\hat H_\mathrm{int}$, the degenerate states with $E=0$ being excluded. We see that the Wigner 
distribution well describes the statistics of most of the intervals. However, the actual distribution shows a tail 
for relatively high values $\Delta E$ that exceeds the Wigner distribution and this excess comprises about 20\% of the energy 
intervals without a clear dependence on $n$. This tail spreads far beyond ${\Delta  E_*}$, therefore many  too high 
values of $\Delta E>0.4\eta $ are not displayed in Fig.~\ref{fig:1}. 

We check next the quantum typicality of the eigenstates of $\hat H_\mathrm{int}$. The considerations based on the ETH 
\cite{Srednicki99} lead to the following estimation of a matrix element of an observable $\hat A$:
\begin{equation} 
\langle \psi _i|\hat A|\psi _j\rangle =\delta _{ij}\bar A(n) +\frac {f(n,E_j-E_i)R_{ij}}{\sqrt{D_n}} .
\label{eq:4} 
\end{equation} 
The diagonal matrix element $\bar A(n)$ depends mainly on $n$ and only weakly on the state $|\psi _j\rangle $. The off-diagonal 
matrix elements consist of $1/\sqrt{D_n}$ as prefactor of an ``envelope function" $f$, which is a smooth function of 
$E_j-E_i$, and $R_{ij}$, which can be considered a random function of $i$ and $j$, 
its sign being positive or negative with equal probability and its absolute value being normalized by 
$D_n^{-1}\sum _j R_{ij}^2 =1$. 
As the observable to be tested we choose the numbers of quanta $N_k$ in the modes $k=1,\, \dots \, ,\, n$. Unlike the basis functions 
$|\{ N_k\} \rangle $, the eigenstates $|\psi _j\rangle $ of the interaction Hamiltonian are not eigenstates of 
$\hat N_k=\hat b_k^\dag \hat b_k$. The matrix elements of interest are 
\begin{equation} 
\langle \psi _i|\hat N_k|\psi _j\rangle =\sum _{\{ N_k\} }\langle \psi _i|\{ N_k \} \rangle N_k\langle \{ N_k \} |\psi _j\rangle  . 
\label{eq:5} 
\end{equation} 

In Fig. \ref{fig:2} we show the mode populations, i.e., diagonal matrix elements $\langle \psi _j|\hat N_k|\psi _j\rangle $ for 
four different eigenstates $|\psi _j\rangle $. The populations are close to the repetition numbers $N_k$ of the number $k$ 
averaged over all possible partitions of $n$, 
\begin{equation} 
\bar N_k =\frac 1{p(n)} \sum _{\{ N_k\} }  N_k , 
\label{eq:6} 
\end{equation} 
as long as they are large enough, $\bar N_k\gtrsim 10^{-1}$. The values of $\bar N_k$ in the same range are well described 
by a Bose-Einstein distribution with zero chemical potential and dimensionless  ``inverse temperature" $\beta $ defined by the 
condition $\sum _{k=1}^n k/[\exp (\beta k)-1]=n$ (for the connection between the Bose-Einstein statistics and partitions of integers see,  
e.g., Refs. \cite{Auluck1946,Rovenchak}).  

\begin{figure}[t] 
\vspace*{1mm} 
\includegraphics[width=\columnwidth]{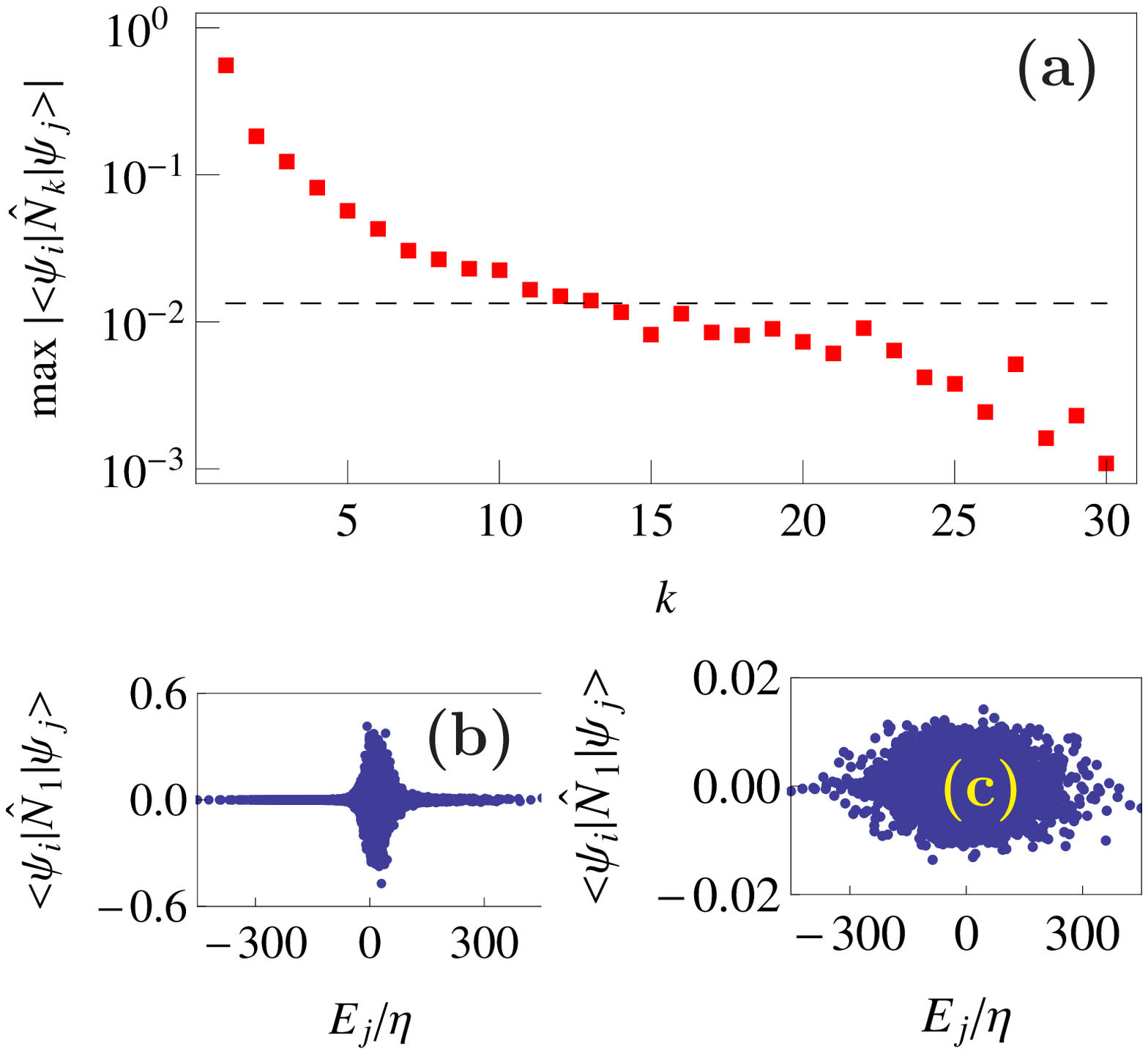} 
\begin{caption} 
{(Color online) Matrix elements $\langle \psi _i|\hat N_k|\psi _j \rangle $ for $i=3181$ ($E_i= 95.349\eta $), $j\neq i$, $n=30$. 
(a) Maximal absolute value of the off-diagonal matrix elements as a function of the mode number (squares). Dashed line: 
$1/\sqrt{D_{30}}=1/\sqrt{5604}$. Off-diagonal matrix elements for all state numbers for (b) $k=1$  and 
(c) $k=11$; the states $j$ are labelled by their energies and ordered by increasing $E_j$. } 
\label{fig:3} 
\end{caption} 
\end{figure}

The properties of off-diagonal matrix elements of $\hat N_k$ 
are illustrated in Fig.~\ref{fig:3} for a state, which is expected to be typical, because 
its energy is neither too large nor too small. We see that $\hat N_1$ couples the states in a narrow band of energies. In contrast, 
off-diagonal matrix elements for $\hat N_{11}$ are uniformly distributed over almost the whole range of states with $j\neq i$. 

Finally, we apply the results of the exact diagonalization of the interaction operator to the solution of a dynamical problem. 
We consider the decay of a phonon in 1D at zero temperature, i.e., we assume that initially, at the time $t=0$ only a single mode is 
populated: $N_k(0)=\delta _{kn}$. In what follows, we work in the interaction representation and set $\hat { b}_k = 
\exp({-i\omega _kt}) \hat {\tilde b}_k$. 
The initial quantum state is $|\Psi (0)\rangle =\hat {\tilde b}_n^\dag |\mathrm{vac}\rangle $, where $|\mathrm{vac}\rangle $ denotes 
the vacuum state of the phonons. The further evolution is given by $|\Psi (t)\rangle =\exp (-i\hat H_\mathrm{int}t/\hbar )
|\Psi (0)\rangle $, and we want to calculate the survival probability $P_n(t) = |\langle \Psi (t)|\Psi (0)\rangle |^2$ in the 
$n$th mode for $t>0$. 

We are aware that this problem is a very idealized toy model, since thermal fluctuations in 1D systems always play an important role 
at experimentally attainable temperatures \cite{OU}, but this problem is a good starting point for demonstrating the agreement 
of the results of our model with those of the field theory and to stress their non-trivial feature (non-exponential decay law). 
The problem of decay of a phonon in 1D at finite temperatures corresponding to the classical limit of the Bose-Einstein 
statistics \cite{Andreev}  will be a subject of our future work. It requires the use of an IPM with a very high dimension, that makes 
full diagonalization of the interaction operator a challenging task. Instead, we can invoke the sparsity of the matrix 
$\langle \{ N_k^\prime \} |\hat H_\mathrm{int}| \{ N_k \} \rangle $ [the number of its non-zero elements is 
$\sim \frac 12 n^2p(n)$] and find the evolution of the system's state of the system using the modern methods for sparse systems of 
differential equations \cite{sparse}. 

From the field-theoretical point of view, the survival probability 
\begin{equation} 
P_n(t)=|G_n^\mathrm{R} (t)|^2 
\label{eq:7}  
\end{equation} 
can be expressed through the 
retarded Green's function $G_n^\mathrm{R}(t) =\frac i{2\pi }\int _{-\infty }^\infty d\omega \, e^{-i\omega t}\tilde G_n^\mathrm{R}(\omega )$ 
of a phonon in vacuum, where  
$\tilde G_n^\mathrm{R}(\omega )=[{\omega -\tilde \Sigma _n (\omega ) +i0^+}]^{-1}$. 
 
A renormalization that replaces bare Green's functions by dressed ones in the standard perturbative expression for the self-energy 
leads to the expression \cite{Samokhin} 
\begin{equation} 
\tilde \Sigma _n(\omega )= \left( \frac \eta \hbar \right) ^2 \int _0^{n/2} dk\, 
\frac {k(n-k)n}{\omega -\tilde \Sigma _k (\omega )-\tilde \Sigma _{n-k} (\omega ) +i0^+} .
\label{eq:8} 
\end{equation} 
A similar equation has been derived by Andreev in the finite-temperature case \cite{Andreev}. 
The off-shell contribution to $\tilde \Sigma _n$ (i.e., coupling to other IPMs in our case) is neglected in Eq.~(\ref{eq:8}) 
\cite{Andreev,Samokhin}. 
We replaced in Eq. (\ref{eq:8}) summation over discrete set of mode numbers by integration, as a traditional approximation in 
field theory, valid for $n\gg 1$.  

In the previous works \cite{Andreev,Samokhin} equations like Eq. (\ref{eq:8}) were solved by setting $\omega \rightarrow 0$. 
In this limit, Eq. (\ref{eq:8}) can be solved exactly, yielding $\tilde \Sigma _n (0) = -i \varsigma \eta n^{2}/\hbar $ with 
$\varsigma = \left \{ \int _0^{1/2} dq\, q(1-q)/[q^{2}+(1-q)^{2}] \right \} ^{1/2} =[(\pi -2)/8]^{1/2} $ and, hence, 
$P_n(t) = \exp ( -2\varsigma \eta n^2t/\hbar )$. However, we show that this approach provides only a correct dependence   
of the typical decay time on $\eta $ and $n$, but the dependence of $\tilde \Sigma _n$ on $\omega $ becomes essential and the 
single-pole approximation for $\tilde G_n^\mathrm{R}(\omega )$ breaks down, therefore the phonon relaxation is non-exponential.

The large frequencies correspond to short times. Therefore, in order to investigate the initial stage of the phonon decay, we 
will expand $\tilde \Sigma _n(\omega )$ and $\tilde G_n^\mathrm{R}(\omega )$ in series in negative powers of $\omega $ that 
converge for $\omega \gtrsim n^2 \eta /\hbar $.  One can see from Eq. (\ref{eq:8}) that the self-energy can be 
expressed as $\tilde \Sigma _n(\omega )=(\omega +i0^+)\Xi (\sigma _n )$, where $\Xi $ is a universal function and its argument 
\begin{equation} 
\sigma _n =\left[ \frac {\eta n^2}{-i\hbar (\omega +i0^+)}\right] ^2
\label{eq:9} 
\end{equation} 
depends on $n$. By changing the integration variable to $q=k/n$ we transform Eq. (\ref{eq:8}) to 
\begin{equation} 
\Xi (\sigma _n) =\sigma _n \int _0 ^{1/2}dq\, 
\frac {q(1-q)}{1+\Xi \left( q^4\sigma _n\right) +\Xi \big( (1-q)^4\sigma _n\big) } .
\label{eq:10} 
\end{equation} 
We make an \textit{ansatz} $\Xi (\sigma _n) =\sum _{l=1}^\infty \Xi _l\sigma _n^l$ and also expand the r.h.s. of Eq. (\ref{eq:10}) 
in powers of $\sigma _n$. The expansion coefficients $\Xi _l$ are determined from comparison of prefactors in front of 
$\sigma _n^l$ on both sides of Eq. (\ref{eq:10}). We obtain $\Xi _1=\frac 1{12}$ and $\Xi _l$ for $l>1$ can be expressed through 
$\Xi _1, \, \dots \, ,\, \Xi _{l-1}$ in a recursive way. 

Knowing $\Xi _l$, one can expand Green's function in series: 
\begin{eqnarray} 
\tilde G_n^\mathrm{R}(\omega )&=& \left \{ (\omega +i0^+) \left[ 1 + 
\Xi \left( -\frac {\eta ^2n^4}{\hbar ^2(\omega +i0^+)^2}\right) \right] \right \} ^{-1}    \nonumber \\ &=& 
\frac 1{\omega +i0^+}\left \{ 1 + \sum _{l=1}^\infty (-1)^l \xi_l \left[ \frac {\eta n^2}{\hbar (\omega +i0^+)}\right] ^{2l}\right \} .
\nonumber \\ && \qquad \label{eq:11} 
\end{eqnarray} 
The inverse Fourier transform yields  an expression
\begin{equation} 
G_n^\mathrm{R}(t) = 1+\sum _{l=1}^\infty \xi _l \frac {(\eta n^2t/\hbar )^{2l}}{(2l)!}
\label{eq:12} 
\end{equation}
that holds for not too large $t$ and is to be substituted into Eq. (\ref{eq:7}). 
The numerically observed convergence of the series allows us to restrict ourselves to a moderate (up to 20) number of 
terms in expansions for $\Xi $ and $G_n^\mathrm{R}$. The first six values of $\xi _l$ are given in Table~\ref{tab:I}.

\begin{figure}[t]  
\vspace*{1mm}  
\includegraphics[width=\columnwidth]{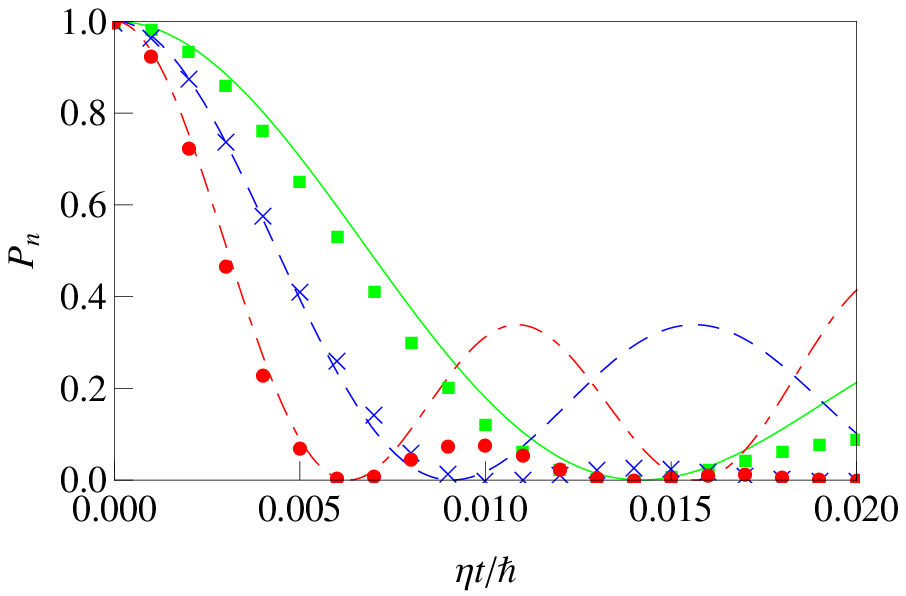} 
\begin{caption}
{(Color online) Survival probability as a function of dimensionless time $\eta t/\hbar $ by two approaches: Eq. (\ref{eq:13}) 
(symbols) and Eqs.~(\ref{eq:7},\,\ref{eq:12}) (lines); $n=20$ (squares, solid line, green), 25 (crosses, dashed line, blue), and 
30 (circles, dot-dashed line, red).  }
\label{fig:4} 
\end{caption}
\end{figure} 

\begin{table}[h] 
\caption{\label{tab:I} Expansion coefficients in Eqs. (\ref{eq:11},\,\ref{eq:12}).}
\begin{ruledtabular}
\begin{tabular}{clclcl}
$l$ & $\qquad ~~\xi _l$ & $l$ & $\qquad ~~\xi _l$ & $l$ & $\qquad ~~\xi _l$ \\
\hline 
1 & $-0.083\,333\,33$ & 3 & $-0.000\,993\,07$ & 5 & $-0.000\,012\,80$   \\
2 & ~~0.008\,928\,57  & 4 & ~~0.000\,112\,24  & 6 & ~~0.000\,001\,47
\\
\end{tabular}
\end{ruledtabular}
\end{table}

On the other hand, the exact  diagonalization of the interaction Hamiltonian allows us to directly compute 
\begin{equation} 
P_n(t) = \bigg{|} \sum _{j=1}^{D_n} e^{-iE_jt/\hbar } |\langle \psi _j |\Psi (0)\rangle |^2 \bigg{|} ^2. 
\label{eq:13}
\end{equation} 
Fig. \ref{fig:4} demonstrates a good agreement between Eq.~(\ref{eq:13}) and the field-theoretical approach 
[Eqs.~(\ref{eq:7},\,\ref{eq:12})] for $\eta n^2t/\hbar \lesssim 5$.  $P_n$ decreases practically to 0 on this time scale, a 
partial revival occurs on longer times.  

Our analysis of the non-exponential decay of a phonon should be distinguished from that of particle tunneling \cite{Garcia}. 
In the latter case, the Green's function is derived by means of single-particle quantum scattering theory, where the boundary conditions for 
outgoing wave are specified. We try to connect the field-theoretical approach to the exact diagonalization of a Hamiltonian 
that models the low-energy effective theory of a many-body 1D system (dynamics of phonons). We hope that our approach to 
non-exponential decay via expansion like Eqs. (\ref{eq:11},\,\ref{eq:12}) can be extended to a broad class of problems. 

The author thanks M. Gluza, A. Gottlieb,  and J. Schmiedmayer for helpful discussions. This work was supported by  
the EU via the ERC advanced grant "QuantumRelax" and by the Wiener Wissenschafts und Technologie Fonds (WWTF) via the grant  	
MA16--066 (project "SEQUEX"). 

\bibliography{references_partition}

\end{document}